\documentstyle[12pt]{article}
\begin{document}
E-print: hep-th/0110011
\vskip1cm
\par
\centerline{\large Supersymmetric Bag Model as a Development}
\centerline{\large of the Witten Superconducting String\footnote{
Talk given at the SUSY'01 conference ( Dubna, June 11-17, 2001) and at the
IX Workshop on High Energy Spin Physics, SPIN-01 (Dubna, August 2-7, 2001).}}

\vskip.5cm
\centerline{A. Ya. Burinskii}
\vskip.25cm
\centerline{\it NSI, Russian Academy of Sciences, e-mail:bur@ibrae.ac.ru}
\vskip.5cm
\begin{quotation}
We consider  particlelike solutions to
supergravity based on the Kerr-Newman BH solution.
The BH singularity is regularized by means of a phase transition to a new
superconducting vacuum state near the core region. We show that this phase
transition can be controlled by gravity in spite of the extreme smallness
of the local gravitational field.

Supersymmetric BPS domain wall model is suggested which can provide this
phase transition and formation the stable charged (dual) superconducting core.
\end{quotation}
\vskip1cm

The simplest consistent Super-Kerr-Newman
BH solution \cite{SBH} was constructed on the base of the
( broken ) Ferrara-Nieuvenhuisen N=2 Einstein-Maxwell D=4 supergravity.
Since source (or singularity) of this solution is covered by BH horizon, the
matter chiral fields of supergravity are not involved at all.
However, for the large angular momentum corresponding to spinning particles
the Kerr horizons are absent, and there appears a naked singularity.
It can be regularized by a matter source \cite{Bag} built of the nontrivial
chiral (Higgs) fields.

   One of the approaches to regularization of the particlelike BH solutions
is based on the old idea of the replacement of singularity by a "semiclosed
world", internal space-time of a constant curvature (M. Markov, 1965;
I. Dymnikova \cite{Dym}).
\footnote{The Dirac classical electron model (generalization of this model
to the charged and rotating bubble in gravity was given by L\`opez, 1984,
see ref. in \cite{Bag}), as well as the bag models could also be included
in this class when one assumes that regularization is provided by a flat
core region.}

We consider development of these models leading to a  non-perturbative
soliton-like solution to supergravity and assuming that the external field
is the Kerr-Newman black hole solution and the core is described by a domain
wall bubble based on the chiral fields of a supersymmetric field model.
\footnote{Close analogy of the BH and domain wall solutions in supergravity
was mentioned in \cite{CvSol}}.

The Kerr-Schild class of metrics
\begin{equation}
g_{\mu \nu} = \eta_{\mu \nu} + 2h K_{\mu}K_{\nu}.
\label{gKS}
\end{equation}
allows one to consider the above regularization
for the rotating and nonrotating, charged and uncharged BH's in unique manner
\cite{Bag}. It allows one to describe the external BH field and the internal
(A)dS region, as well as a smooth interpolating region between them without
especial matching conditions, by using one smooth
function, $f(r)$, of the Kerr radial coordinate $r$.

Here $\eta$ is an
auxiliary Minkowski metric, $K_\mu$ is a vortex field of the Kerr
principal null congruence, and scalar function $h$ has the form
\footnote{For $a\ne 0$ the Kerr coordinates $r$ and $\theta$ are oblate
spheroidal ones.}
\begin{equation}
h=f(r)/(r^2 +a^2\cos ^2 \theta).
\label{hf}
\end{equation}
 In particular, for the Kerr-Newman BH solution
\begin{equation}
f(r)=f_{ext}(r)=mr-e^2/2,
\label{fKN}
\end{equation}
where $m$ and $e$ are the total mass and charge.
 The transfer to nonrotating case occurs by $a=0$ when the Kerr
congruence turns into a twist-free "hedgehog" configuration, and $r, \theta$
are usual spherical coordinates.
It is important that function $f(r)$ is not affected by this transfer
that allows one to simplify treatment concentrating on the $a=0$ case.

By $a=0$, the regularizing core region of a constant curvature can be
described by $f=f_{int}(r)= \alpha r^4 $, where $\alpha=\Lambda/6 $,
$\Lambda$ is cosmological constant, and energy density in core is
\footnote{In this case, as shows (\ref{hf}),
gravitational singularity is regularized also by $a\ne 0$.}
$\rho= \frac 3 4 \alpha / \pi$ .

A smooth matching of the internal and external metrics is provided by
smooth function $f(r)$ interpolating between $f_{int}$ and $f_{KN}$.
The radial position $r_0$ of the phase transition region can be estimated
as a point of intersection of the plots $f_{int}$ and $f_{ext}$,
\begin{equation}
\frac 4 3 \pi \rho r_0 ^4=mr_0-e^2/2.
\label{r0}
\end{equation}

Analysis shows \cite{Bag} that for charged sources there appears a
thin intermediate shell at $r= r_0$ with a strong tangential stress that
is typical for a domain wall structure.
Dividing this equation on $r_0$ one can recognize here the {\it mass balance
equation}
\begin{equation}
m = M_{int}(r_0) + M_{em}(r_0),
\label{mb}
\end{equation}
where $m$ is total mass, $M_{int}(r_0) $ is ADM mass of core and
$ M_{em}(r_0) = e^2/2r_0 $ is ADM mass of the external e.m. field.
It should be mentioned, that gravitational field is extremely small at $r_0$,
especially as $r_0$ is much more of gravitational radius
\footnote{In particular, if interior is flat ($\rho =0$ ) $r_0= e^2/2m$
-`classical electromagnetic radius'.}
( $r_0/m \sim 10^{42}$). Nevertheless, eq.
(\ref{mb}) shows that {\it phase transition is controlled by
gravity, but nonlocally!}
Note, that $M_{int}$ can be either
positive (that corresponds to dS interior) or negative ( AdS interior ).
As we shall see, supergravity suggests AdS vacua inside the bubble.

As consequence of this treatment we obtain also some demands to the
supergravity matter field model.

i - It has to provide a phase transition between internal
and external vacua.

ii - External vacuum has to be (super)-Kerr-Newman black hole solution
with {\it long range } electromagnetic field and  zero cosmological constant.

iii -Internal vacuum has to be (A)dS space with superconducting properties.

These demands are very restrictive and cannot satisfied in the known
    solitonlike bag, domain wall and bubble models.  Main contradiction is
connected with demands ii) and iii) since in the most of models external
electromagnetic field is short range. An exclusion is the
$U(I) \times \tilde U(I) $ field model which was used by Witten to describe
the cosmic superconducting strings \cite{Wit}. Our suggestion is to use
this field model for description the superconducting baglike configuration.
The model contains two sectors A and B (two Higgs and two gauge fields).
One of the gauge fields (sector A) we set as long range external
electromagnetic field. It acquires mass at the core region where the Higgs
A-field forms a configuration similar to "lumps", Q-balls and other
non-topological solitons \cite{Col}, but with a specifical form of potential.
The other Higgs field (sector B) forms a superconducting bag with
confined inside the bag second gauge field.

     Supersymmetric version of the Witten field model (suggested by
J. Morris \cite{Mor}) has effective Lagrangian of the form
\begin{eqnarray}
L=-2(D^\mu \phi )\overline {( D_\mu  \phi  )}-2(\tilde D^\mu \sigma )
(\overline {\tilde D_\mu \sigma} )-\partial ^\mu Z \partial _\mu
\bar Z \nonumber \\
-\frac 14F^{\mu \nu }F_{\mu \nu }-
\frac 14 F_B ^{\mu \nu }F_{B\mu \nu}-V (\sigma, \phi, Z),
\label{bur-SL}
\end{eqnarray}
where the potential $V$ is determined through
the superpotential $W$ as
\begin{equation}
V= \sum _{i=1} ^5 \vert\partial _i W \vert^2,
\label{SV}
\end{equation}
and the superpotential $W(\Phi^i)$ is a holomorphic function of
the fife complex chiral fields
$\Phi ^i = \lbrace Z,\phi, \bar\phi, \sigma, \bar \sigma \rbrace $,
\begin{equation}
W=\lambda {Z}(\sigma \bar\sigma -\eta ^2) + ( c Z+m ) \phi\bar \phi.
\label{bur-SW}
\end{equation}
\par
In the effective Lagrangian the "bar" is identified with complex
conjugation, so there are really only three independent scalar fields, and the
"new" ( neutral ) fields Z provides the synchronization of the phase
transition.
The supersymmetric vacuum states corresponding to the lowest value of the
potential are determined by the conditions
\begin{eqnarray}
 \partial _i W  =0;
\label{Svac}
\end{eqnarray}
and yield $V=0$.
These equations lead to two supersymmetric vacuum states:
\begin{equation}
I ) \qquad Z=0;\quad \phi=0 ;\quad \vert\sigma\vert=\eta ;\quad W=0,
\label{bur-true}
\end{equation}
we set it for external vacuum; and
\begin{equation}
II ) \qquad Z=-m/c;\quad \sigma=0; \quad \vert \phi \vert =\eta
\sqrt{\lambda/c};\quad W=\lambda m\eta ^2/c,
\label{bur-false}
\end{equation}
we set it as a state inside the bag.
\par

The treatment of the gauge field $A_\mu$ and $B_\mu$ in $B$-sector is similar
in many respects because of the symmetry between  $A$ and $B$ sectors
allowing one to consider the state $\Sigma = \eta$ in outer region as
superconducting one \footnote{The version of dual superconductivity
in B-sector seems the most interesting.}
in respect to the gauge field $B_\mu$.
Field $B_\mu$ acquires the mass $m_B= g \eta $ in outer region, and the
$\tilde U(I)$ gauge symmetry is broken, which provides confinement of the
$B_{\mu}$ field inside the bag.
The bag can also be filled by quantum excitations of fermionic,
or non Abelian fields.
\par
One can check the phase transition in the planar wall approximation
 ( neglecting the gauge fields ). It can be shown that it is a
BPS-saturated
domain wall solution interpolating between supersymmetric vacua I) and II).
Using the Bogomol'nyi
transformation one can represent the energy density as follows
\begin{eqnarray}
\rho &=& T_{00} =\frac{1}{2} \delta _{ij}\lbrack ( \Phi ^i,_z)
(\Phi ^j,_z) + (\frac {\partial W}{\partial \Phi ^i})
(\frac {\partial W}{\partial \Phi ^j})\rbrack \\
 &=&\frac{1}{2} \delta _{ij}\lbrack  \Phi ^i,_z + \frac {\partial W}{\partial
\Phi ^j}\rbrack
\lbrack  \Phi ^j,_z + \frac {\partial W}{\partial\Phi ^i}\rbrack
- \frac{\partial W}{\partial \Phi ^i}\Phi ^i,_z,
\label{trbog}\end{eqnarray}
where the last term is full derivative.
Then, integrating over the wall depth $z$ one obtains for the surface
energy density of the wall
\begin{equation}
\epsilon=\int _0 ^\infty \rho dz =\frac 12 \int\Sigma _i
( \Phi ^i,_z + \frac {\partial W}{\partial\Phi ^i})^2 dz + W(0) -W(\infty).
\label{edens}
\end{equation}
The minimum of energy is achieved when the first-order Bogomol'nyi
equations $\Phi ^i,_z + \frac {\partial W}{\partial\Phi ^i} =0$ are
satisfied, or in terms of $Z, \Phi, \Sigma $
\begin{eqnarray}
 Z^\prime &=& -  \lambda(\Sigma ^2-\eta ^2) - c \Phi ^2, \\
 \Sigma ^\prime &=& -\lambda Z\Sigma, \\
\Phi ^\prime &=& - (cZ +m) \Phi.
\label{bogeq}
\end{eqnarray}
Its value is given by
$\epsilon =W(0)-W(\infty)=\lambda m \eta^2/c$.
Therefore, this domain wall is BPS-saturated solution.
One can see that
the field $Z$, which appears only in the supersymmetric version
of the model, plays an essential role for formation of the phase transition.
\par
The structure of stress-energy tensor contains the typical for domain walls
tangential stress. The non-zero components of the stress-energy tensor have
the form
\begin{eqnarray}
T_{00} & = &-T_{xx} =- T_{yy}=\frac{1}{2}[ \delta _{ij}( \Phi ^i,_z)
(\Phi ^j,_z) + V];\\
T_{zz}& = &\frac{1}{2}[ \delta_{ij}( \Phi ^i,_z)(\Phi ^j,_z) - V] =0.
\label{bur-Tflat}
\end{eqnarray}

The energy of an uncharged bubble forming from the BPS domain wall is
\begin{equation}
E_{0 bubble} = E_{wall} =
4\pi \int _0 ^\infty \rho r^2 dr \approx 4\pi r_0 ^2
\epsilon.
\label{E0tot}
\end{equation}
However, the Tolman mass
$M =\int dx^3 \sqrt{-g}(-T_0^0+T_1^1 +T_2^2 +T_3^3)$,
 taking into account tangential stress of the wall, is
negative
\begin{equation}
M_{Tolm. bubble} = - E_{wall} \approx -4\pi r_0 ^2 \epsilon.
\label{M0tot}
\end{equation}
It shows that the uncharged bubbles are unstable and
form  the time-dependent states \cite{CvSol}.
\par
Charged bubbles have extra
contribution caused by the energy and mass of the external electromagnetic
field
\begin{equation}
E_{e.m.} = M_{e.m.} = \frac{e^2}{2r_0},
\label{EMem}
\end{equation}
and contribution to mass caused by gravitational field of the external
electromagnetic field ( determined by Tolman relation for
the external e.m. field)
\begin{equation}
M_{grav. e.m.} =  E_{e.m.} = \frac{e^2}{2r_0}.
\label{Mgrem}
\end{equation}
As a result the total energy for charged bubble is
\begin{equation}
E_{tot.bubble} = E_{wall} + E_{e.m.} = 4\pi r_0 ^2 \epsilon
+ \frac{e^2}{2r_0},
\label{Etot}
\end{equation}
and the total mass will be
\begin{equation}
M_{tot.bubble} = M_{0 bubble} + M_{e.m.} + M_{grav.e.m.} =
- E_{wall} + 2E_{e.m.} = -4\pi r_0 ^2 \epsilon + \frac{e^2}{r_0}.
\label{Mtot}
\end{equation}
Minimum of the total energy is achieved by
\begin{equation}
r_0=(\frac{e^2}{16\pi \epsilon_{min}})^{1/3},
\label{rmin}
\end{equation}
which yields the following expressions for total mass and energy of
the stationary state
\begin{equation}
M_{tot}^* =E_{tot}^* =\frac {3e^2}{4r_0}.
\label{EMst}
\end{equation}
One sees that the resulting total mass of charged bubble is positive,
however, due to negative contribution of $M_{0 bubble}$ it can be lower
than BPS energy bound of the domain wall forming this bubble.
This is a remarkable property of the bubble models, existence the
`ultra-extreme' states \cite {CvSol} ) gives a hope to overcome
BPS bound and get the ratio $m^2 \ll e^2$ which is
necessary for particle-like models.

 For the rotating Kerr case  $J=ma$ and for $J\sim 1$ one finds out that
parameter $a\sim 1/m$ has Compton size.
Coordinate $r$ is an oblate spheroidal coordinate, and matter is foliated on
the rotating ellipsoidal layers \cite{Bag}. Curvature of space is
concentrated in equatorial plane, near the former singular ring, forming a
stringlike tube. \footnote{For the parameters of electron the phase
transition region represents an oblate rotating disk of Compton size
and thickness $\sim e^2/2m$.}

In supergravity, for strong fields there is also an extra contribution to
stress-energy tensor leading to negative cosmological constant
$\Lambda =-3 k^4 e^{k^2K}\vert W \vert ^2$ which can yield AdS space-time
for the bag interior. The treatment shows that:
\begin{itemize}
\item
 in spite of the extreme smallness of the local gravitational
field supergravity can control the position of phase transition
at unexpectedly large distances;
\item
core of the Kerr spinning particle has a disklike shape, and one can
expect a sensitivity of differential sections for polarized spinning
particles depending on the direction of polarization.
\end{itemize}

The considered here supersymmetric model is more complicated than the
traditionally used domain wall models \cite {CvSol}, and it demonstrates
some new properties.
One of the peculiarities of this model is the presence of gauge fields
which, as it was shown in thin wall approximation, allow one to stabilize
 bubble to a finite size. Second peculiarity is the presence of a few
chiral superfields that can give a nontrivial sense to K\"ahler metric
$ K^{i\bar j}$ of the supergravity field models. One can expect that
extra degrees of freedom of the K\"ahler metric can play a role
in formation of the bent domain wall configurations.
On the other hand, the connected with superconductivity chiral fields
acquire a nontrivial geometrical interpretation in the
Seiberg--Witten theory and in the Landau -- Ginzburg
theory where the chiral superfields  refer to the moduli of the internal
Calabi--Yau spaces \cite{GVW}. It gives an interesting link to higher
dimensions with an alternative look on compactification.

We are thankful to  M. Cveti\v c,  S. Hildebrandt, S. Manayenkov, A. Efremov,
O. Teryaev and A. Wipf for useful discussions.

\end{document}